# Pressure Induced Enhancement of Superconductivity in $LaRu_2P_2$

Baoxuan Li, Pengchao Lu, Jianzhong Liu, Jian Sun, Sheng Li, Xiyu Zhu & Hai-Hu Wen*

National Laboratory of Solid State Microstructures and Department of Physics, Collaborative Innovation Center of Advanced Microstructures, Nanjing University, Nanjing 210093, China

Correspondence and requests for materials should be addressed to HHW (email: hhwen@nju.edu.cn).

Abstract:

**To explore new superconductors beyond the copper-based and iron-based systems is very important. The Ru element locates just below the Fe in the periodic table and behaves like the Fe in many ways. One of the common thread to induce high temperature superconductivity is to introduce moderate correlation into the system[1-3]. In this paper, we report the significant enhancement of superconducting transition temperature from 3.84K to 5.77K by using a pressure only of 1.74 GPa in $LaRu_2P_2$ which has an iso-structure of the iron-based 122 superconductors. The *ab-initio* calculation shows that the superconductivity in $LaRu_2P_2$ at ambient pressure can be explained by the McMillan's theory with strong electron-phonon coupling. However, it is difficult to interpret the significant enhancement of $T_c$ versus pressure within this picture. Detailed analysis of the pressure induced evolution of resistivity and upper**

**critical field $H_{c2}(T)$ reveals that the increases of $T_c$ with pressure may be accompanied by the involvement of extra electronic correlation effect. This suggests that the Ru-based system has some commonality as the Fe-based superconductors.**

The transition metal compounds exhibit extremely interesting and rich physics due to the close energies concerning the charge, spin and orbital dynamics[4,5]. Usually the band width *t* is narrow in the *3d* transition metals, like Cu, Fe etc., therefore, compared with the electron repulsion energy (the so-called Hubbard U, about 1-5 eV), *t/U* is small and the correlation is quite strong. One of the consequence of this strong correlation is the formation of the magnetic state and possibly the local Cooper pairing[2,6,7,8,9,10]. In some *4d* transition metal compounds, the larger bandwidth makes the charge carriers more mobile against the strong Coulomb effect, yielding many exotic properties. The Ru element is a typical *4d* transition metal element which locates just below the Fe element in the periodic table and exhibits many appealing features. For example, *p*-wave superconductivity may exist in the $Sr_2RuO_4$ compound[11-13]. The $LaRu_2P_2$ has long been known to be a superconductor[14,15] with $T_c \approx 3.8$ K. Interestingly this compound has a similar structure as the parent phase $BaFe_2As_2$ of many iron-based superconductors in the 122 family[16-19]. Therefore it is very interesting to investigate what is the superconducting mechanism in this Ru-based 122 system. Here we report the pressure induced enhancement of superconductivity in a $LaRu_2P_2$ single crystal.

In Figure 1a we show the temperature dependence of resistivity of the $LaRu_2P_2$ single crystal at ambient pressure and 1.74GPa. Details about the growth and characterization of the crystal are given in Methods and Supplementary Information. It is clear that the residual resistivity ratio RRR=ρ(300K)/ρ(5K)= 27.0 is quite large, this indicates the cleanness of the crystal. The inset of Fig. 1a shows the magnetic susceptibility near the superconducting transition at a field of 20 Oe with the zero-field-cooled (ZFC) and field-cooled (FC) mode. Fig.1b presents the temperature dependence of resistivity under different pressures from ambient to 2.25 GPa, one can see that the superconducting transition temperature $T_c$ is clearly increased from 3.84K to 5.77K with the pressure increased to about 1.74 GPa, then $T_c$ drops down slightly with further increase of pressure. Here the $T_c$ value was determined at the temperature with 50% of the normal state resistivity $\rho_n$. The enhancement ratio of $T_c$ versus pressure, i.e. $dT_c/dp$, is about 1.11 K/GPa. This value is quite big and very surprise to us, since it is comparable to that in some unconventional superconductors[20,21].

At a pressure higher than 1.74 GPa, the $T_c$ is getting lower. The $T_c$ value versus pressure shows a dome like $T_c$-p phase diagram, as shown in Fig.2a. This behavior is different from that in the previously reported AC susceptibility measurements with hydrostatic pressures[22], which reveals also an enhanced $T_c$, but superconductivity suddenly disappears above 2.1GPa. In order to understand the superconductivity mechanism, we performed *ab-initio* calculations for the electron-phonon coupling in the frame work of density functional perturbation theory[23], details of the calculations were provided in the Method section and Supplementary Information. Based on the McMillan theory for strong electron-phonon coupling, the superconducting transition temperature can be estimated as:

$$T_c = \frac{\omega_{\log}}{1.45} \exp\left[ -\frac{1.04(1+\lambda_{e-ph})}{\lambda_{e-ph} - \mu \times \left(1 + 0.62\lambda_{e-ph}\right)} \right] \qquad (1)$$

Here $\omega_{\log}$ is the maximum phonon frequency and $\lambda_{e\text{-}ph}$ is the electron-phonon coupling constant, $\mu$ is the Coulomb screening constant. Using the values of $\omega_{\log}$ and $\lambda_{e\text{-}ph}$ from the calculation for the pristine sample and taking $\mu$=0.12, we get a superconducting transition temperature $T_c$ = 3.9 K. This is very close to our experimental value $T_c$ = 3.84K. This may indicate that the superconductivity in the sample at ambient pressure is induced by the electron-phonon coupling, which is consistent with the conclusion drawn previously[24]. In order to understand the pressurized effect, we also did the calculations under pressures up to 5 GPa. With the calculated values of $\omega_{log}$ and $\lambda_{e\text{-}ph}$ under different pressures, as shown in Fig. 2b, we do find a non-monotonic change of $\omega_{log}$ and $\lambda_{e\text{-}ph}$ at a pressure of about 2 GPa. However, if we input all these calculated quantities into the McMillan's formula (eq.1), we only get a slight enhancement of $T_c$, but much weaker than that observed in the experiment. This suggests that the enhancement of superconducting transition temperature here cannot be interpreted as purely due to the phonon mediated pairing.

In order to get a deeper insight of the pressure induced enhancement of superconductivity, we take a look at the normal state resistivity under a pressure. The experimental data and the fitting results are presented in Fig.3a as symbols and solid lines, respectively. We fit the data of ρ vs. T in the temperature range from just above $T_c$ to about 30K with the general formula $\rho = a + bT^c$. Here $a$ represents the residual resistivity due to the impurity scattering. According to the Matthiessens's rule, the composed resistivity can be written as:

$$\rho = a + bT^c = \frac{m^*}{ne^2}\frac{1}{\tau_{imp}} + \frac{m^*}{ne^2}\frac{1}{\tau} \qquad (2)$$

Here $m^*$ is the effective mass when the quasiparticles are moving cross the lattice, $n$ is the effective charge carrier density, $1/\tau_{imp}$ is the impurity scattering rate which is positively related to the impurity density $n_{imp}$, $1/\tau$ is the scattering rate with the lattice. From the fitting results shown in Fig.3b, c, d, one can see that, $a = m^* / ne^2\tau_{imp}$ increases with pressure and almost doubles at a pressure of 1.97 GPa, then it turns to flatten off at higher pressures. The pre-factor $b$ is quite complicated, it is related not only to the effective mass $m^*$, but also to the electron-phonon coupling in a complex way. Therefore the increase of residual resistivity or $a$ may be explained as the increase of $m^*$, since $m^* = m_0(1 + \lambda_{e-ph} + \lambda_{e-boson})$ with $m_0$ the bare mass of the electron, $\lambda_{e-boson}$ is the extra electron-boson coupling strength in addition to the electron-phonon coupling. In this simple argument, we can reasonably expect that $1 + \lambda_{e-ph} + \lambda_{e-boson}$ increases for about two times when the pressure is increased from zero to 1.74 GPa. In Fig.3d, we present the pressure dependence of the power exponent *c* which decreases from about 2.8 to 2.0 with the pressure enhanced from zero to 1.74 GPa. This is consistent with the picture that extra electron-boson coupling sets in and induces the crossover from the electron-phonon coupling to a moderate correlation effect. In the simple phonon scattering picture, a power law of $1/\tau \propto T^n$ (n=3~5) was predicted in the low temperature region[25]. With the involvement of correlation effect, the power exponent *n* will be lowered down to 2 in the Fermi liquid picture[26]. Therefore the evolution of *a*, *b* and *c* with pressure can be self-consistently explained. This also explains why the superconducting transition temperature is increased much faster than that predicted by the picture with simple phonon mediated pairing.

In order to give support to the picture mentioned above, we measured the temperature dependence of the upper critical field of the sample under ambient and a pressure of 1.94 GPa. In Fig.4a and 4b, we show the resistive transitions of the sample under these two states at different magnetic fields. We determined the superconducting transition temperature using the 50%$\rho_n$ criterion and present the data in Fig.4c. It is clear that not only the $T_c$ value is increased, the slope -d$H_{c2}$/dT changes from about 248 Oe/K at ambient pressure to about 601Oe/K at 1.94GPa. According to the Ginzburg-Landau theory, near $T_c$ it was estimated that[27] $-dH_{c2}(T)/dT \propto N_{eff}^{\alpha}$ with $N_{eff} = N_F^0(1+\lambda_{e-ph}+\lambda_{e-boson})$, the effective density of states (DOS) with the total electron-boson coupling constant $\lambda_{e-ph}+\lambda_{e-boson}$, α=0.5~1, where $N_F^0$ is the bare DOS at the Fermi energy. The increase of $-dH_{c2}/dT$ with pressure is very consistent with our previous conclusion that the enhancement of superconductivity is actually induced by the involvement of some extra electron boson coupling which makes the system change from electron-phonon dominated to moderate correlation governed Cooper pairing.

In the following we try to get some insights based on our ab-initio calculations. First, we relax the structure under different pressures and find a clear structural change upon pressure. As shown in Fig. 5a, the change of the Ru-P bond length within the Ru-P conducting layer exhibits an unusual step-wise feature at pressure of 2.0 - 3.5 GPa. This is accompanied with a slight closing of the P-Ru-P angle in the $RuP_4$ tetrahedron and the slope for the change of the angle also varied a bit in the same pressure range. We can also find an obvious change on the electronic structures level. As we can see from Fig. 5b and the Supplementary Information, four bands depicted with red, blue, pink and orange construct a complicated Fermi surface together. One band mostly contributed from Ru 4d ($d_{xz}+d_{yz}$) and P 3pz (red) moves upward with pressure and across the Fermi level at about 3.0 GPa, meanwhile, another band mainly consist of Ru 4d ($d_{xz}+d_{yz}$) orbital (blue) moves downwards and the small wave-like feature near the N point also across the Fermi level at the same pressure range. This blue band opens a small tunnel in the Fermi surface after 3 GPa, as shown in the Supplementary Information. *This movement slightly reduces the slope of the bands across the Fermi level, which may be very essential to enhance the effective electron mass and induce a moderate correlation effect.* From the electronic DOS in Fig. 5c, one can also see a clear variation from 2 to 3.5 GPa. The change of the Ru atom is mainly originated from its $d_{xz}+d_{yz}$ orbitals while change of the P atom is mostly come from its $3p_z$ orbital. Fig. 5d shows the phonon spectra, phonon linewidth, phonon density of states, Eliashberg function $\alpha^2F$, electron-phonon coupling (EPC) constant $\lambda$ calculated for 2 GPa. It seems that a phonon mode near the N point have relatively large phonon linewidth and good contribution to the EP coupling. As shown in the Supplementary Information, this mode slightly goes soft with pressure and reach the lowest frequency at 2.5-3 GPa and goes harder afterwards. The EP coupling constant reach a maximum value of 0.80 at about 2 GPa, which is somehow larger than that of the iron-arsenide system. As mentioned before, with all these refined structural parameters, the significant enhancement of $T_c$ versus pressure in $LaRu_2P_2$ cannot be interpreted purely by the electron-phonon coupling, extra electron-boson coupling may have been involved in the formation of superconducting pairing. The present work strongly suggests that $LaRu_2P_2$ may have some commonalities as the iron based superconductors[28,29] in which the spin fluctuations are supposed to be the major play in forming the Copper pairs. This will stimulate future studies on the interplay of superconductivity and magnetism in the Ru-based systems.

## Methods

### I. Sample growth and measurement techniques

The single crystals $LaRu_2P_2$ were grown by flux method, using polycrystalline $LaRu_2P_2$ as precursor. The starting materials La metal scraps (99%), Ru powder (99.9%, Alfa Aesar), and phosphor powder(99.9%, Alfa Aesar) were weighed in stoichiometric ratio and mixed together, put into an alumina crucible. All these procedures were done in a glove box filled with Ar atmosphere. The crucible was sealed in an evacuated quartz ampule and kept at 1000℃ for 24 hours. The sintered $LaRu_2P_2$ powder was mixed with Sn flux in the molar ratio 1:40 and loaded into an alumina crucible, which was sealed in an evacuated quartz ampule and kept inside a PID controlled furnace box. It was raised to 1100℃ at a rate of 60℃/h and maintained for 4 days, then the temperature was reduced down very slowly to 750℃ at a rate of 1.5℃/h. The Sn flux was centrifuged out at 750℃ before cooling down to room temperature. Some flux sticking to the crystal surface was dissolved in an aqueous solution of hydrochloric acid.

X-ray diffraction (XRD) measurements were performed on a Bruker D8 Advanced diffractometer with the Cu-K$\alpha$ radiation. DC magnetization measurements were carried out with a SQUID-VSM-7T (Quantum Design). Measurements of resistivity under pressure were performed up to 2.3 GPa on a physical property measurement system (PPMS-16T, Quantum Design) by using a HPC-33 Piston type pressure cell with the Quantum Design dc resistivity and ac transport options. For the resistive measurements, silver leads with a diameter of 50$\mu$m were glued to the $LaRu_2P_2$ single crystal in a standard four-probe method by using silver epoxy, and the sample was immersed in the pressure transmitting medium(Daphne 7373) in a Teflon capsule with a diameter of 4 mm. Hydrostatic pressure was generated by a BeCu/NiCrAl clamped piston-cylinder cell. The pressure upon the sample was calibrated with the shift in $T_c$ of a high purity Sn sample by measuring the temperature dependence of resistivity.

### II. Ab-initio calculation

First-principles calculations are performed using the Quantum-ESPRESSO code[30],

ultrasoft pseudopotentials with the Perdew-Burke-Ernzerhof (PBE)[31] Generalized Gradient Approximation (GGA) density functionals are employed, phonon and electron-phonon coupling calculations are carried out within density functional perturbation theory (DFPT) framework[23]. The cutoffs are 80 Ry for the wave functions and 800 Ry for the charge density. The self-consistent calculations are performed over a 12 x 12 x 12 k-point grid. A denser 24 x 24 x 24 grid is used for evaluating an accurate EP interaction matrix. Dynamical matrices and the electron-phonon coupling were calculated on a 4x4x4 q-point mesh.

**Acknowledgements:** This work was supported by the Ministry of Science and Technology of China (973 projects: 2011CBA00102, 2012CB821403, 2010CB923002, 2015CB921202), the National Natural Science Foundation of China (Grant No. 51372112, 11190023, 11034011,

11190020), PAPD and the Fundamental Research Funds for the Central Universities. Part of the calculations were performed on the IBM Blade cluster system in the High Performance Computing Center of Nanjing University.

**Author Contributions:** The samples were fabricated by B.X.L, with the help of X.Y.Z. The transport measurements were done by B.X.L. and J.Z.L. The ab-initio calculations were done by J.S and P.C.L. H-H.W. coordinated the whole work and wrote the manuscript, which was supplemented by other co-authors. All authors have discussed the results and the interpretation.

**Author Information:** The authors declare no competing financial interests.

## Figures and Captions

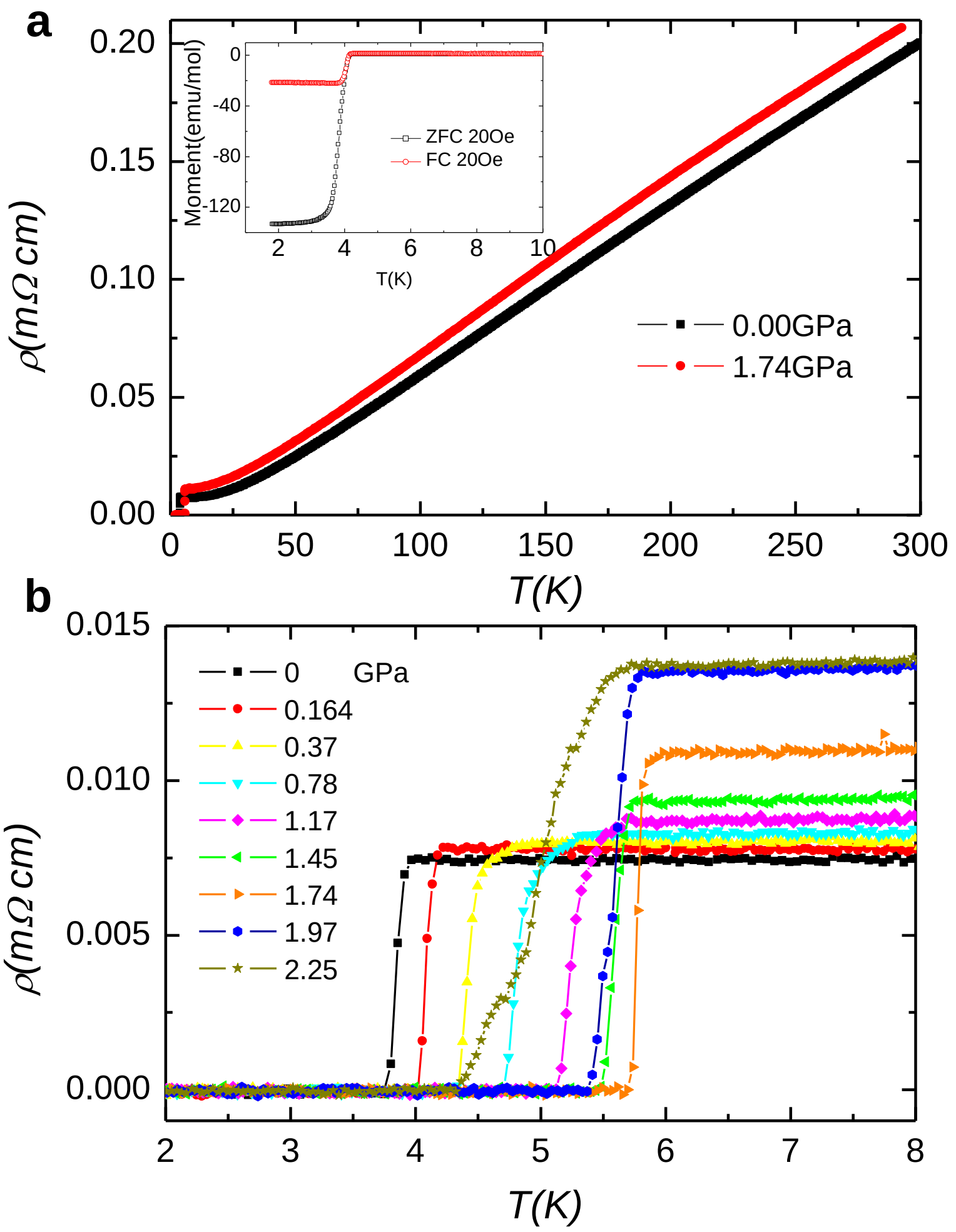

**Figure 1 Pressure induced evolution of superconducting transitions. a,** Temperature dependence of electrical resistivity for the $LaRu_2P_2$ single crystal in the temperature range 2K to 300K as measured under ambient pressure and 1.74GPa, shown by the black square and red circle symbols, respectively. The inset shows the temperature dependence of dc magnetic susceptibility of the sample as measured at an applied magnetic field of 20Oe at ambient pressure. Both the magnetic susceptibility measured in zero-field-cooled (ZFC) and field-cooled (FC) modes are shown. **b,** Temperature dependence of electrical resistivity for the $LaRu_2P_2$ single crystal under various pressures from ambient to 2.25 GPa.

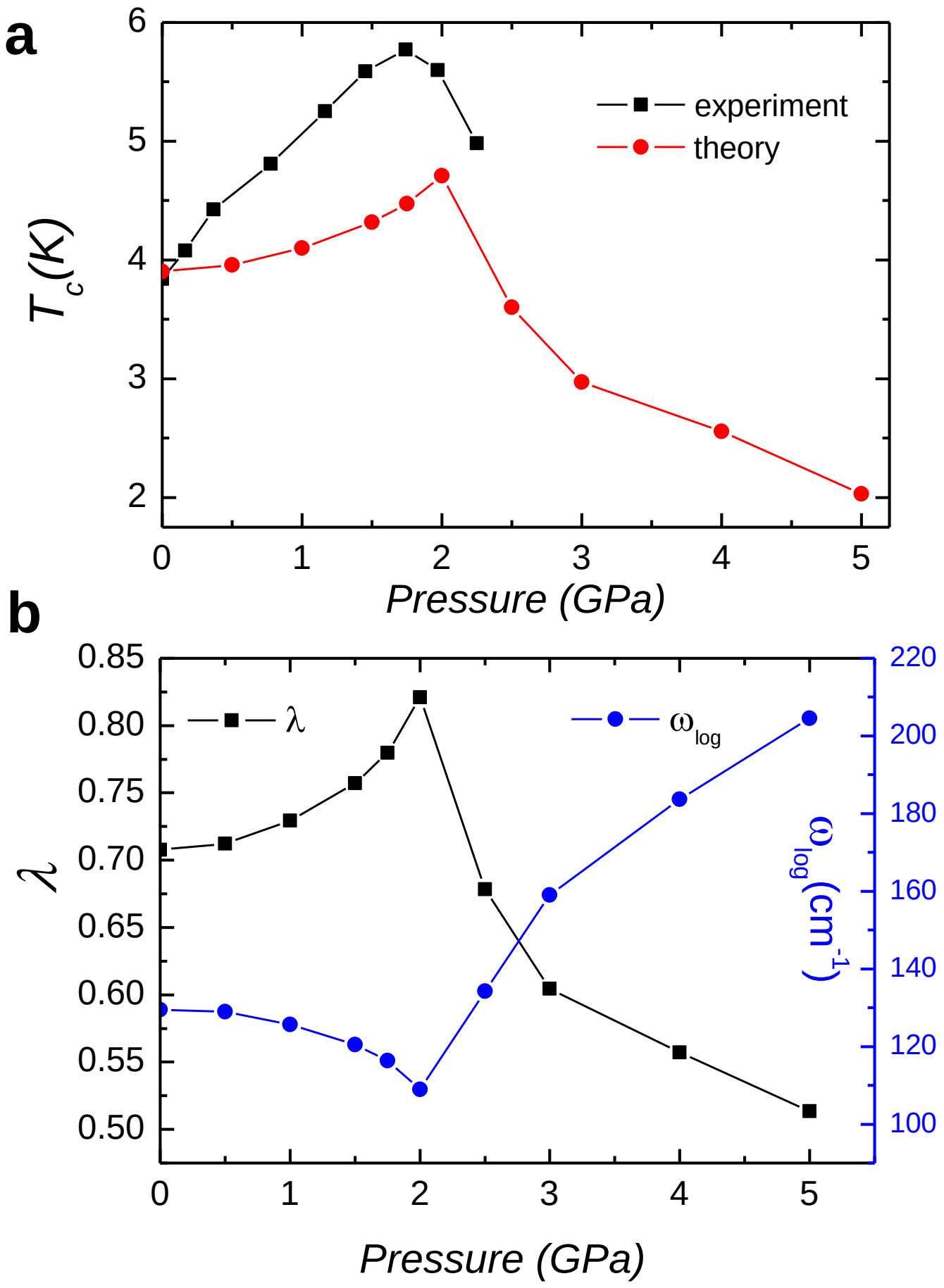

**Figure 2 Pressure enhanced superconductivity and theoretical calculations. a,** Transition temperature $T_c$ versus pressure for the $LaRu_2P_2$ single crystal obtained by the electrical resistivity measurement (black squares). Theoretical calculation results (red circles) are presented in the same graph for comparison, which is based on the McMillan's theory for the case of strong electron-phonon coupling. **b,** First principle calculation values of the maximum phonon frequency $\omega_{log}$ and the electron-phonon coupling constant $\lambda_{e\text{-}ph}$ under different pressures. The calculation is based on the Density Functional Theory, using ultrasoft pseudopotentials, Generalized Gradient Approximation (PBE) functionals.

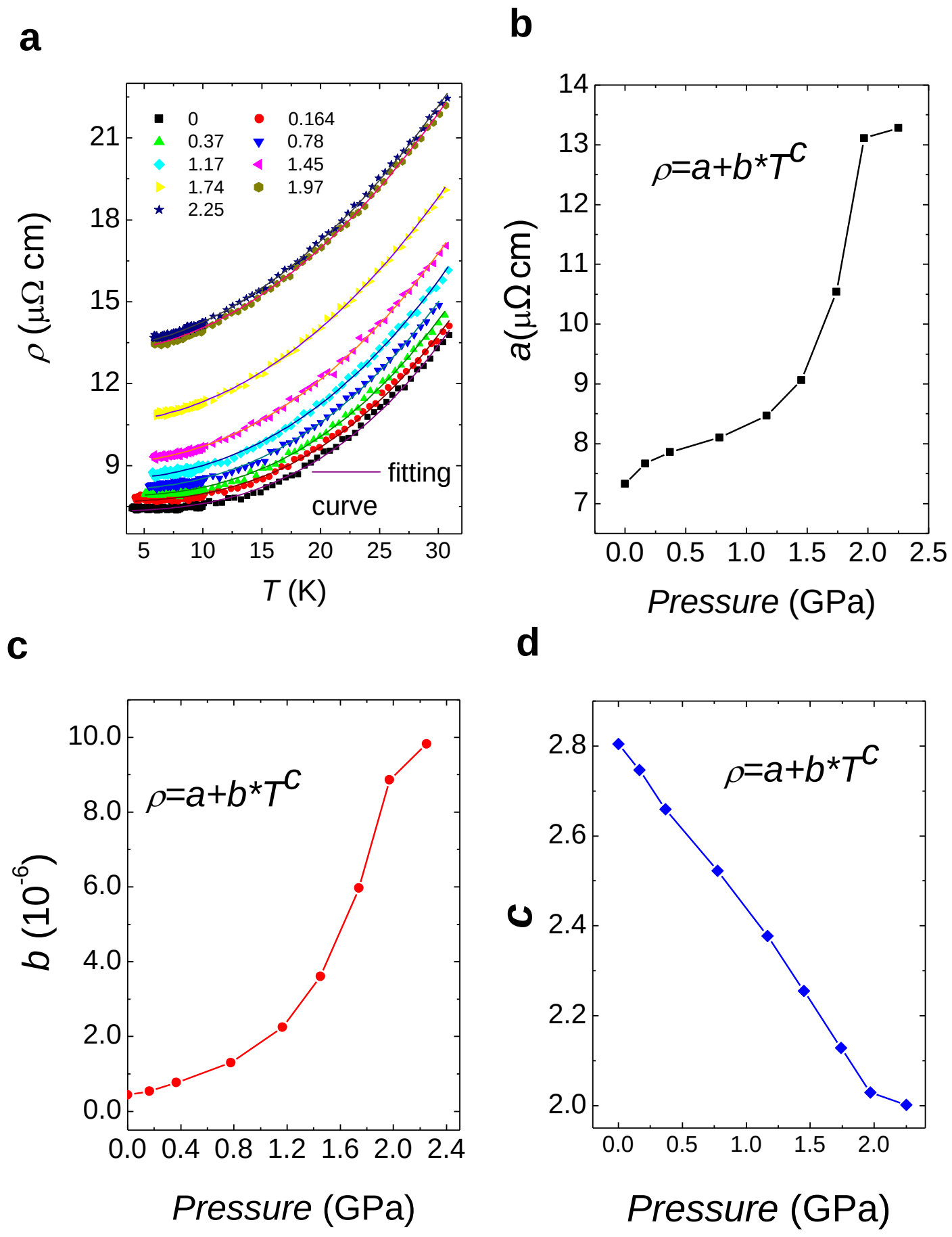

**Figure 3 Fitting to the resistivity data under different pressures. a,** The electrical resistivity data and the fitting results under various pressures as shown by symbols and solid lines, respectively. The data of resistivity versus temperature is fitted in the temperature range of just above $T_c$ to about 30K with the general formula $\rho = a + bT^c$, here $a$ represents the residual resistivity due to the impurity scattering. The obtained coefficients $a$, $b$, $c$ under different pressures are presented in **b**, **c**, and **d**, respectively.

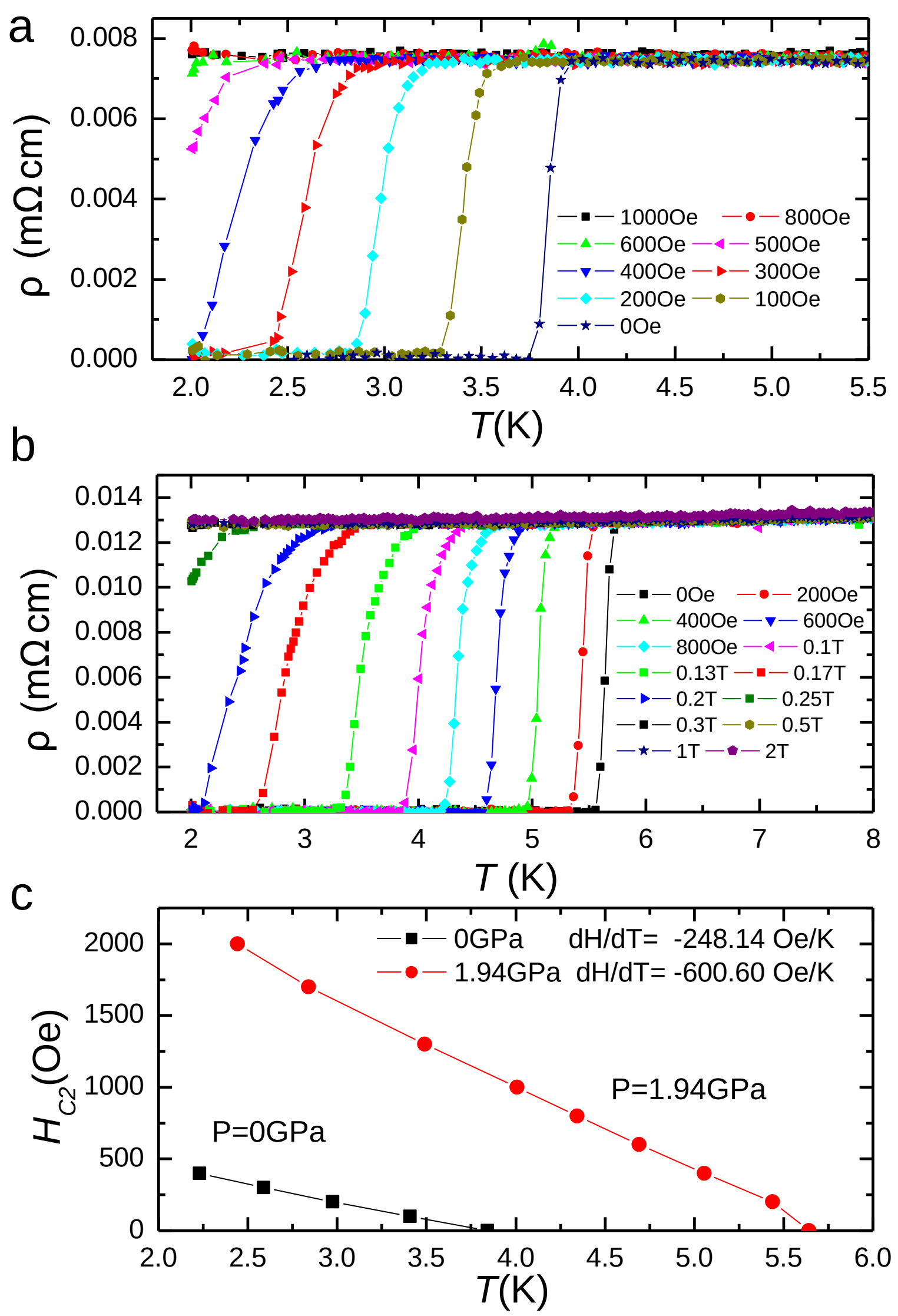

**Figure 4 Resistive transitions and upper critical fields under ambient pressure and 1.94 GPa.** **a,** Temperature dependence of electrical resistivity for $LaRu_2P_2$ in the temperature range 2K to 4.5K at ambient pressure measured under various magnetic fields perpendicular to the basal plane. **b,** Temperature dependence of electrical resistivity of the sample in the temperature range 2K to 8K with a pressure of 1.94GPa measured under various magnetic fields perpendicular to the plane. **c,** Temperature dependence of the upper critical field for the $LaRu_2P_2$ single crystal under ambient pressure and with a pressure of 1.94 GPa, as shown by black squares and red circles, respectively. The transition temperature is obtained from Fig 4**a** and 4**b**, using the 50%$\rho_n$ criterion.

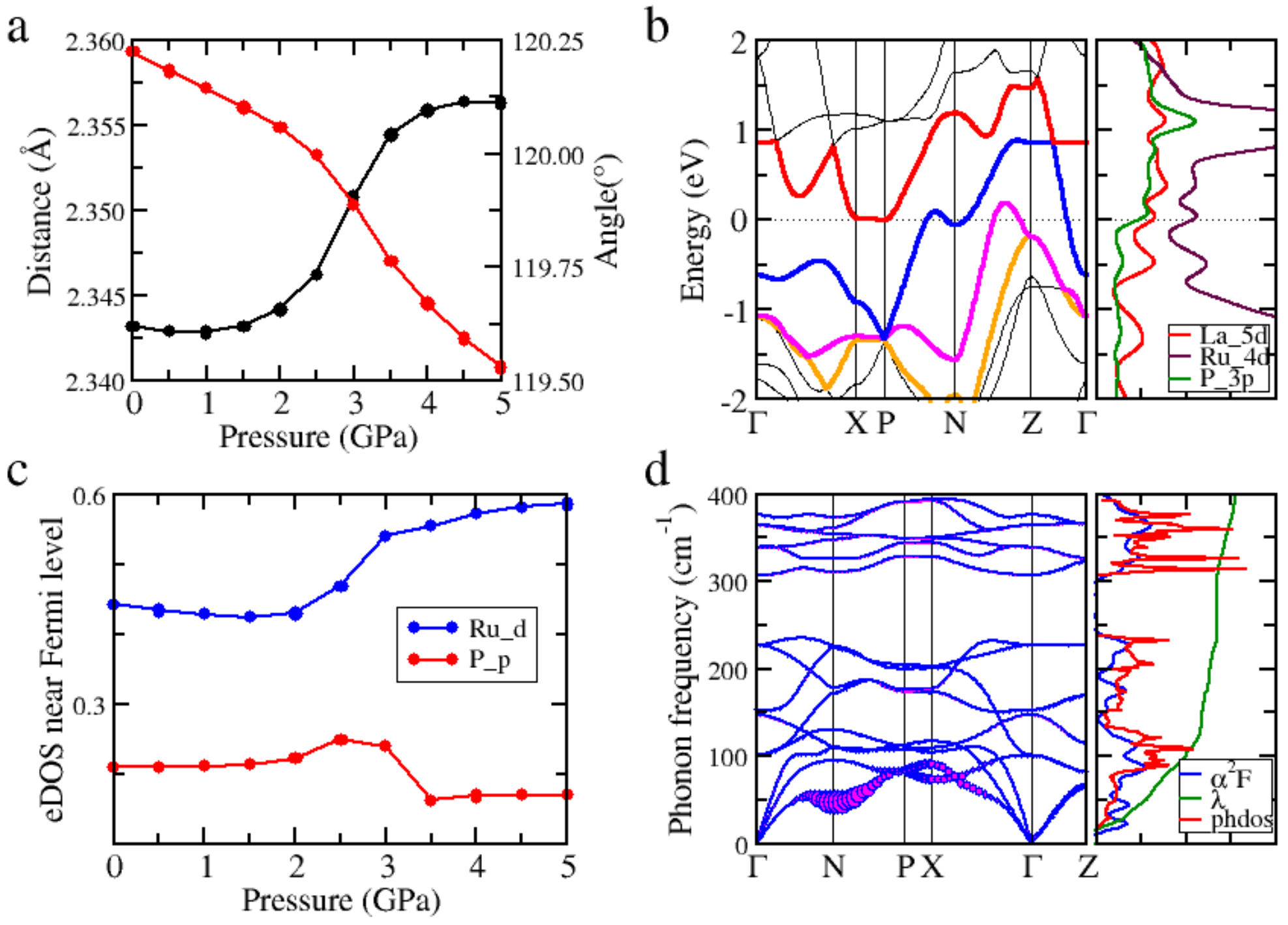

**Figure 5 Theoretical calculations under different pressures. a,** Calculated values of Ru-P bond length (red line) and P-Ru-P angle (black line) in the $RuP_4$ tetrahedron vs. pressure; An obvious enhancement of the Ru-P bond length (and reducing of the P-Ru-P angle) can be found at the pressure from 2 to 3.5 GPa. **b,** The electronic band structure and DOS of $LaRu_2P_2$ (I4/mmm) with calculated lattice parameters at 3 GPa; four bands close to Fermi level are marked with different colors. The band lying on Fermi level between X and P (black) at 3GPa are mainly contributed by P 3p($p_z$) and Ru 4d($d_{xz}$+$d_{yz}$) orbit. Panel on the right shows three partial DOS which make main contributions to the total DOS. **c,** Calculated electronic DOS near Fermi level vs pressure, presents an increasing of Ru 4d orbit and a dropping of P 3p orbit, corresponding to **a**. **d,** Calculated phonon dispersions at 2 GPa; the size of the bubble represents the electron-phonon interaction magnitude; phonon DOS and the integral value of electron-phonon coefficient λ are also shown on the right panel.

# Supplementary Information for

# Pressure Induced Enhancement of Superconductivity in $LaRu_2P_2$

Baoxuan Li, Pengchao Lu, Jianzhong Liu, Jian Sun, Sheng Li, Xiyu Zhu & Hai-Hu Wen*

## I. Characterization of the $LaRu_2P_2$ crystal

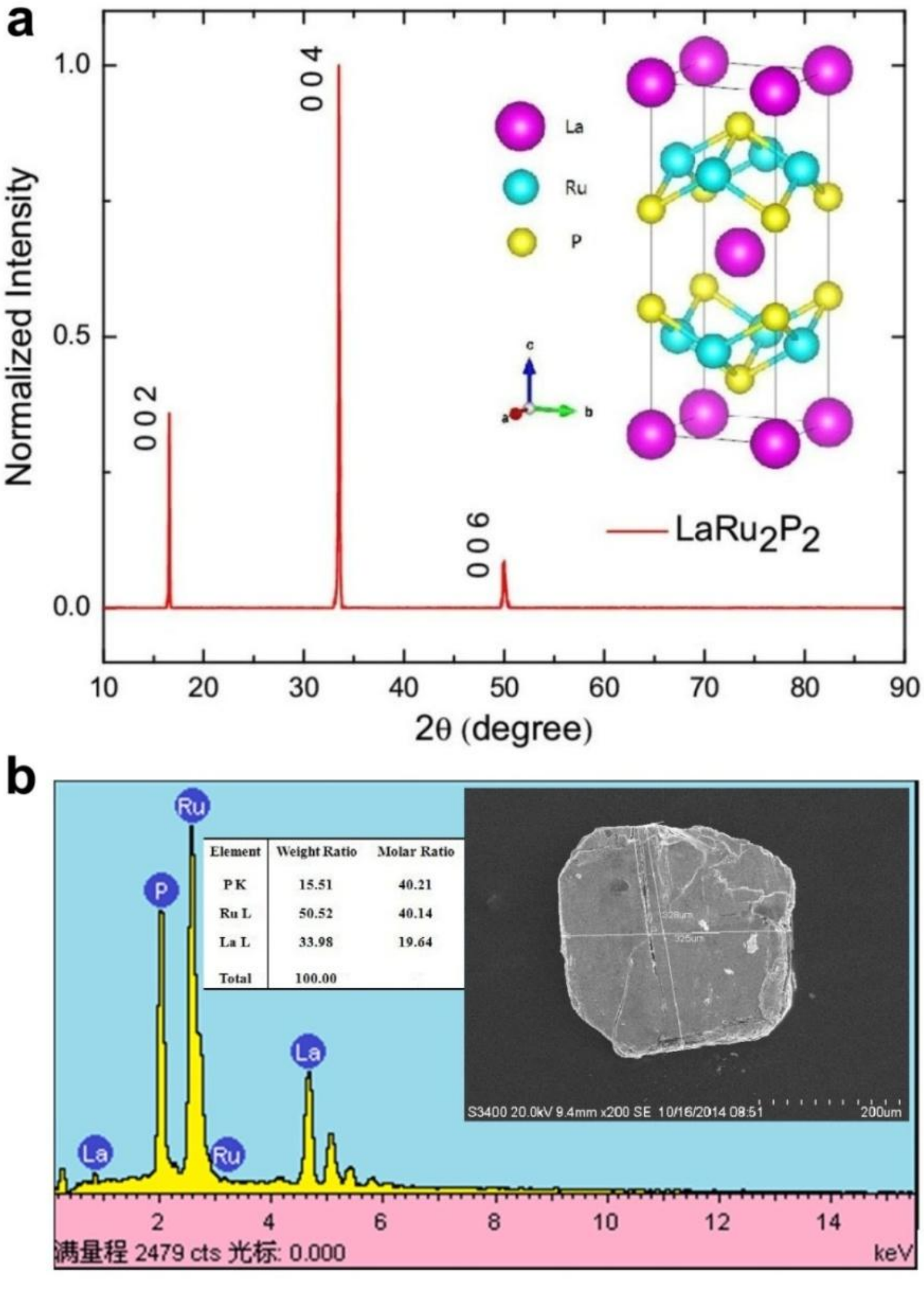

**Figure S1 Characterization of the $LaRu_2P_2$ crystal. a,** X-ray diffraction pattern for a $LaRu_2P_2$ single crystal. The inset gives the schematic show of the atomic structure of $LaRu_2P_2$, which adopts the tetragonal $ThCr_2Si_2$-type structure (I4/mmm space group). **b,** Energy Dispersive X-ray microanalysis pattern for $LaRu_2P_2$ single crystal. The inset shows the SEM photograph of the crystal with dimensions of about $325\times325\times40\mu m^3$.

In Fig.S1a we present the X-ray diffraction pattern for the $LaRu_2P_2$ single crystal. It is clear that only the (0 0 l) reflections can be detected, and we can obtain the c-axis lattice constant $c$ = 10.690Å from these peaks. Energy dispersive X-ray spectrum (EDS) measurements were performed at an accelerating voltage of 20kV and working distance of 10 millimeters by a scanning electron microscope (Hitachi Co.,Ltd.). Fig.S1b shows one set of the EDS result on a $LaRu_2P_2$ single crystal, the atomic ratio is very close to La:Ru:P = 1:2:2, probably with some La vacancies, as is shown in the table in Fig.S1b.

## II. Ab-initio calculation on the pressure induced change of structural and band structures

1. Change of the lattice constants with pressure

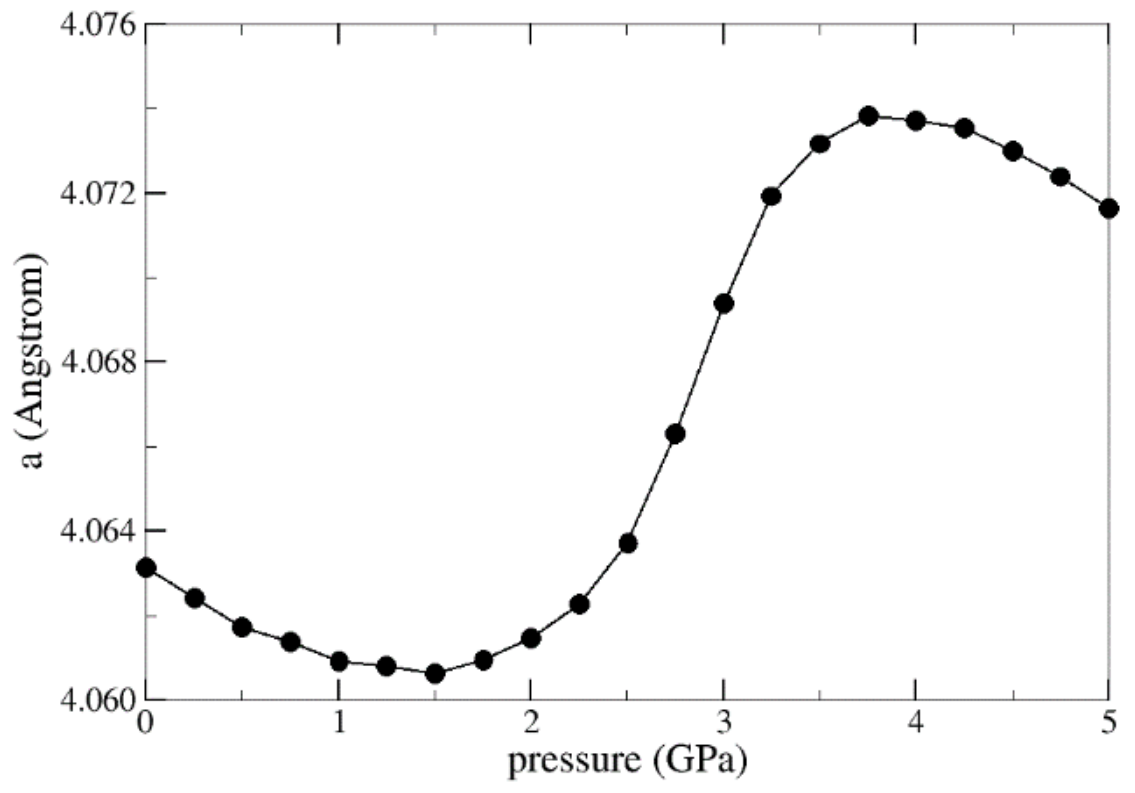

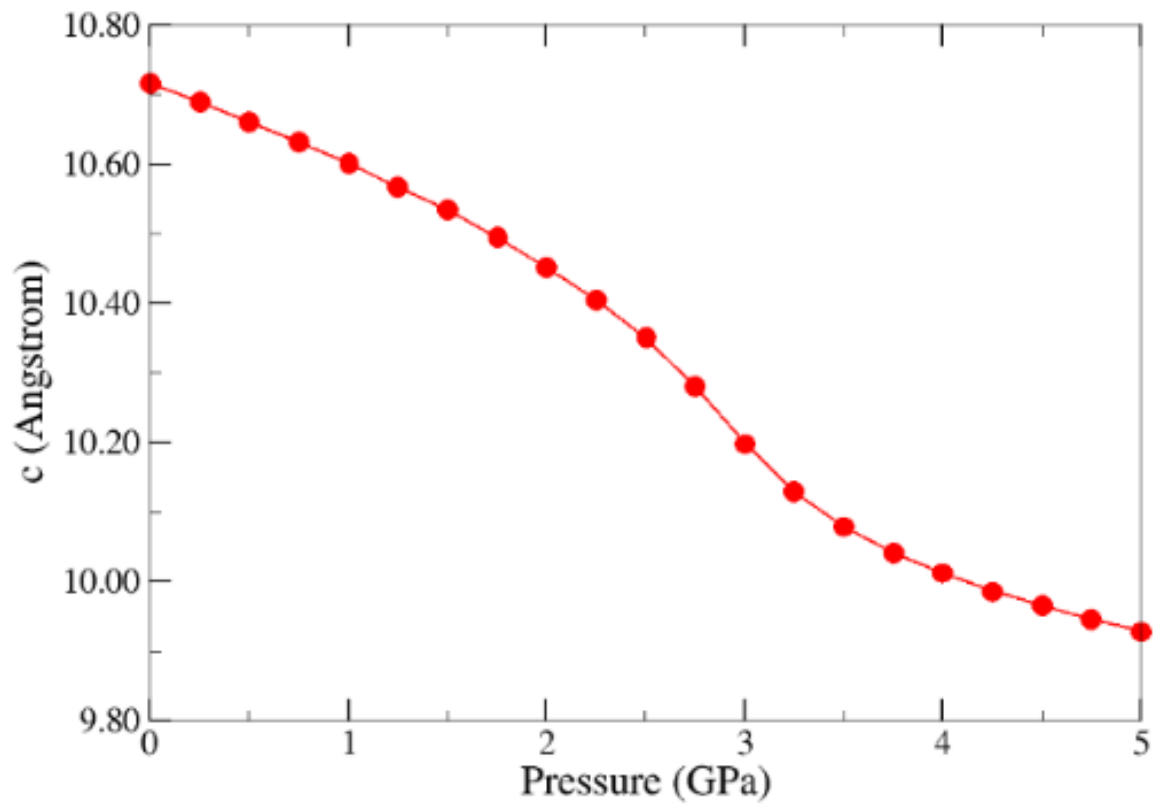

2. Change of the band structures with pressure

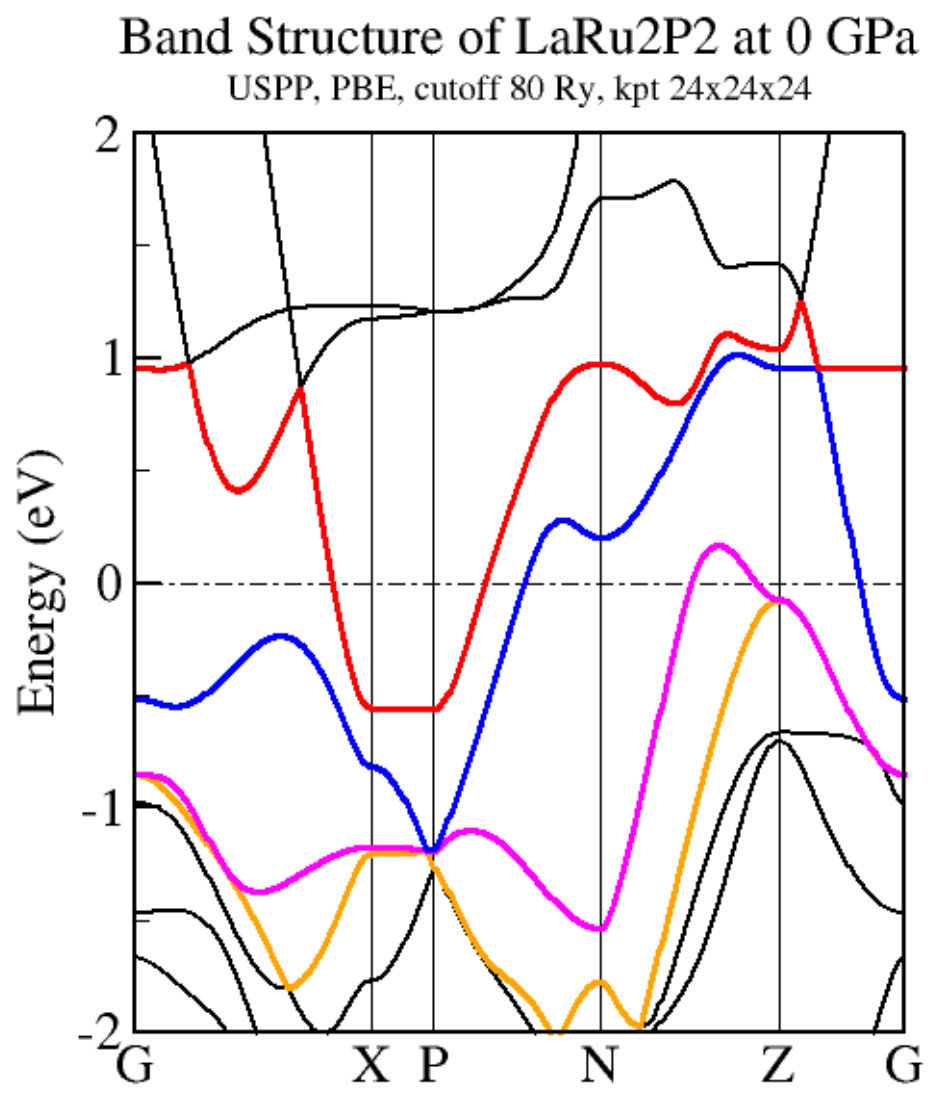

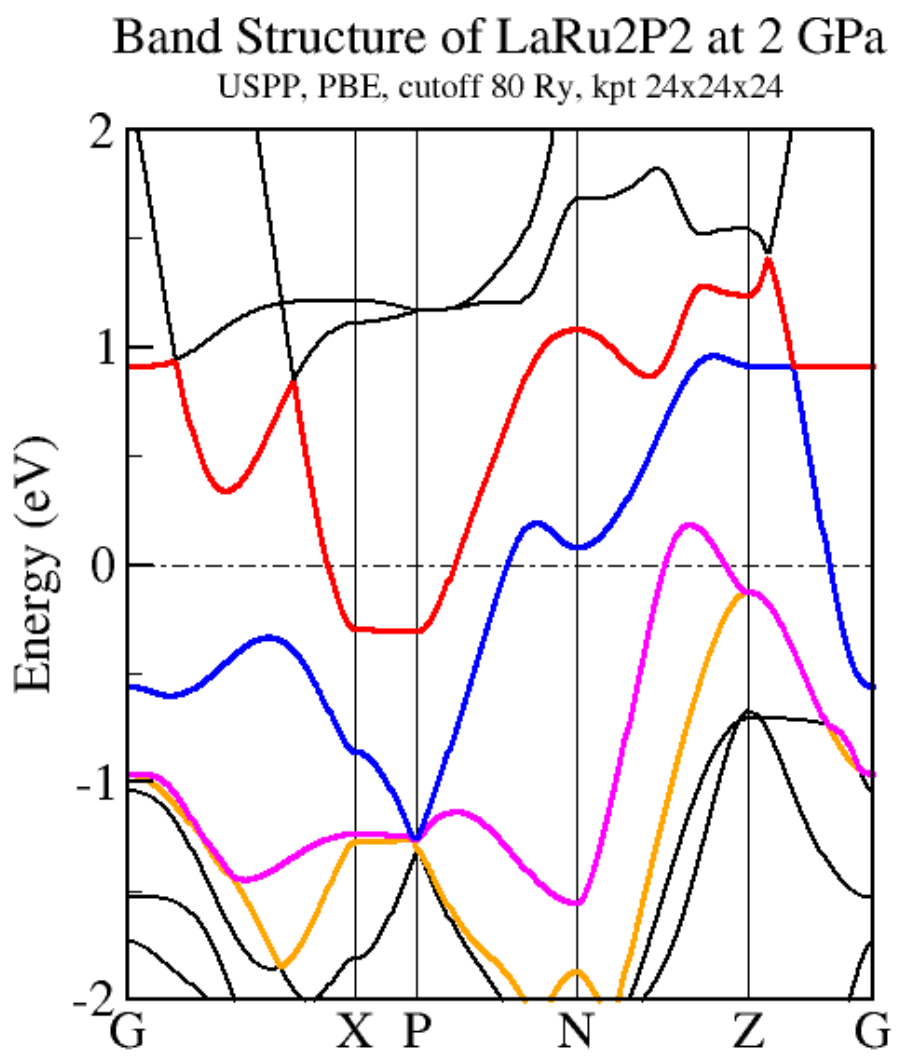

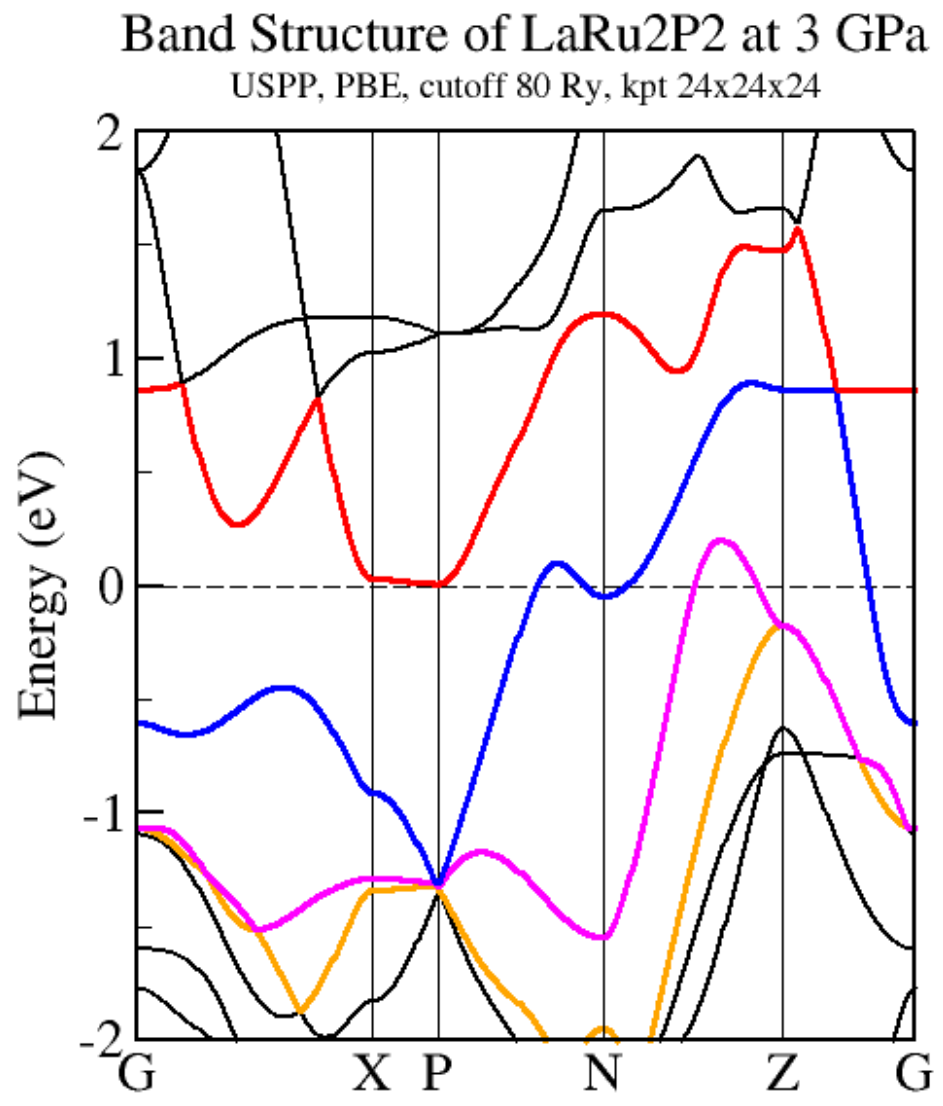

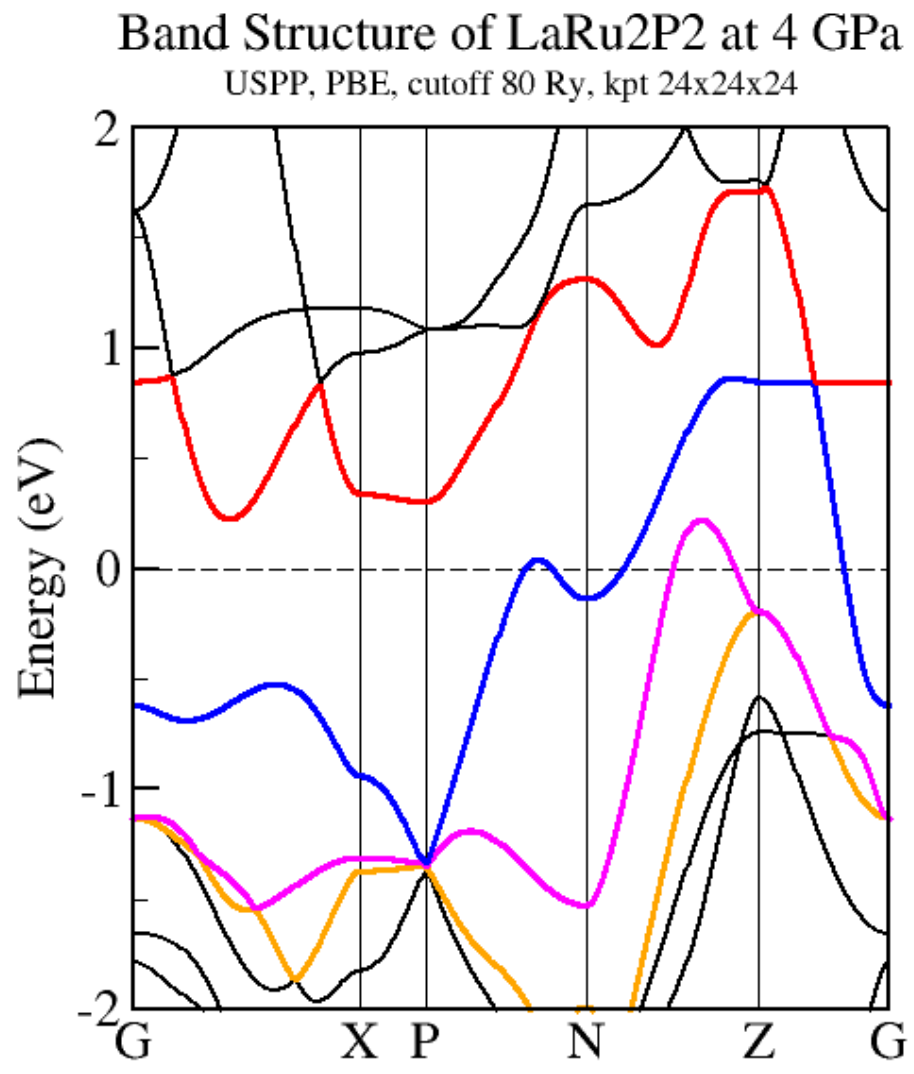

3. Change of the DOS with pressure

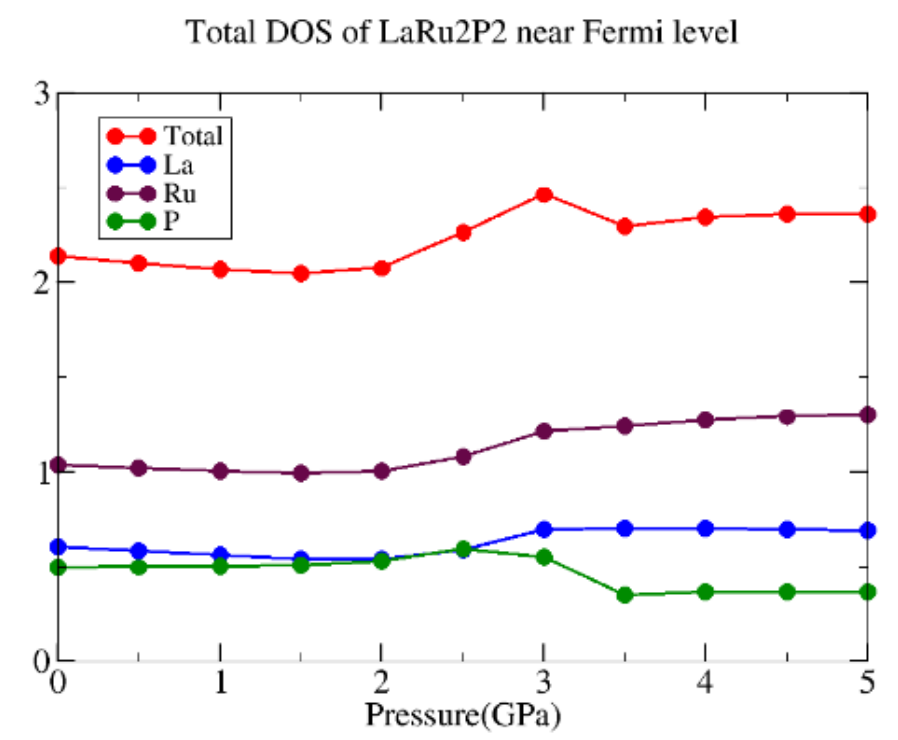

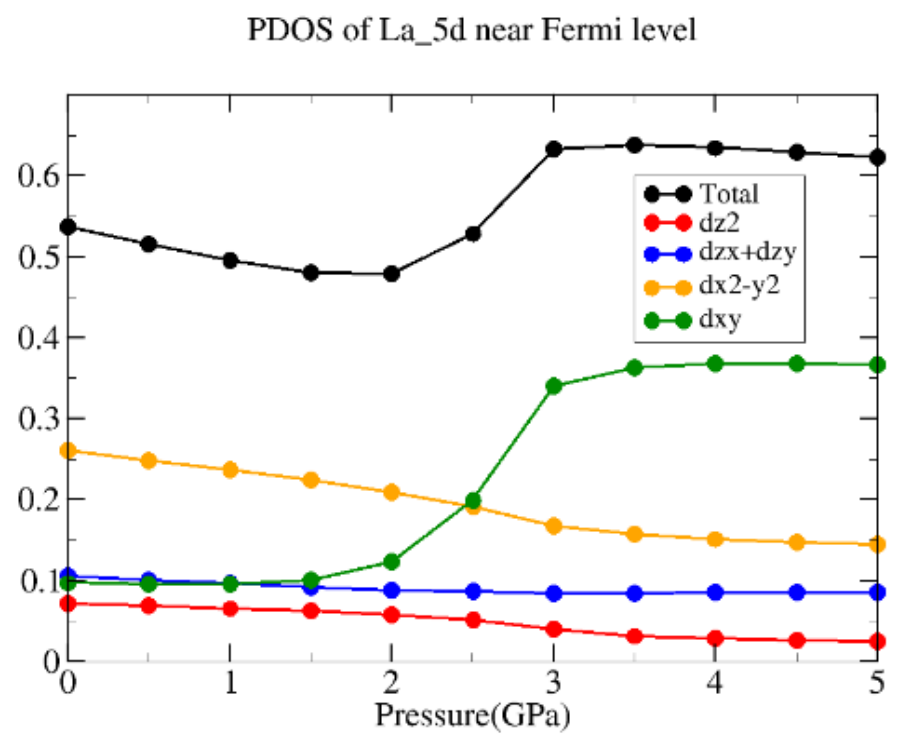

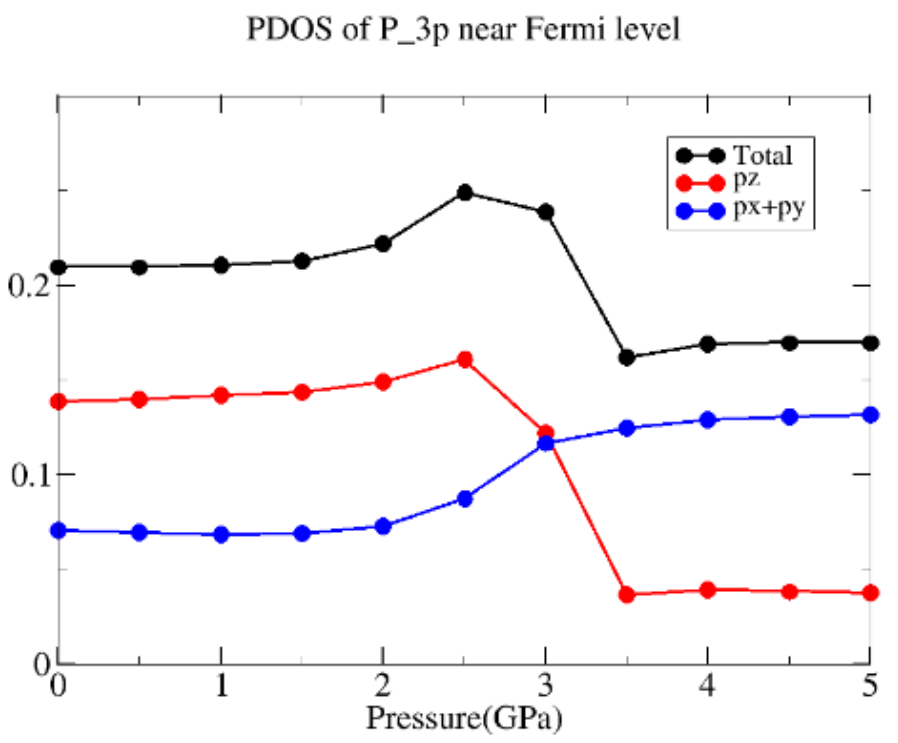

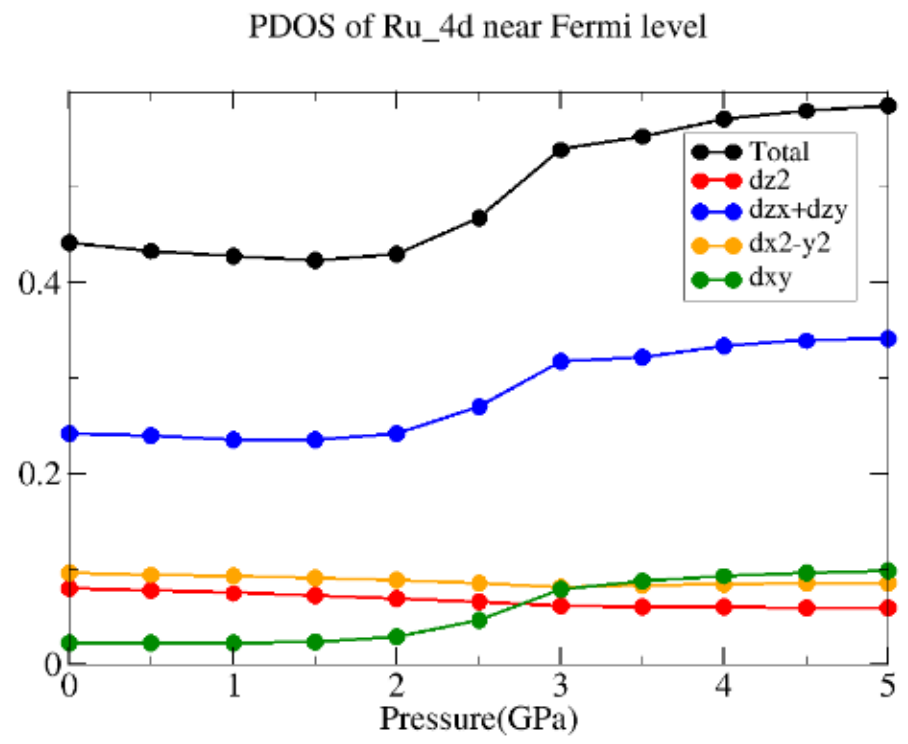

4. Change of the Fermi surface with pressure

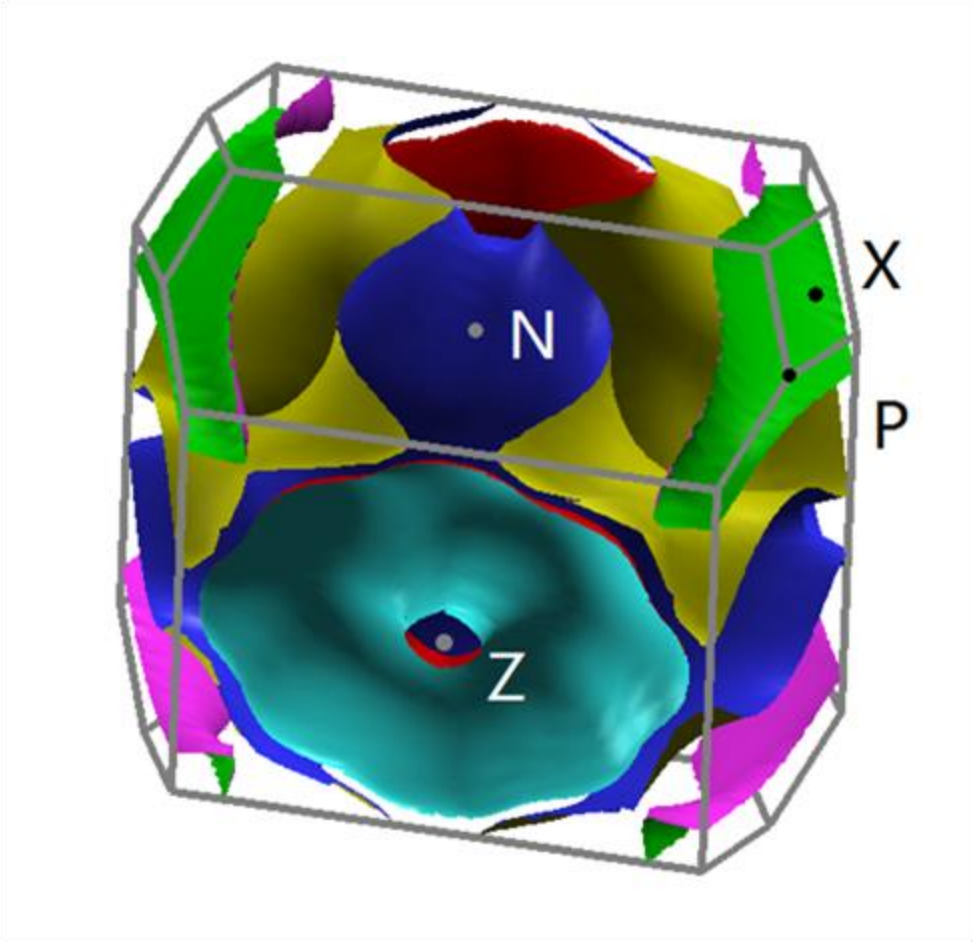

Fermi surface at 2GPa

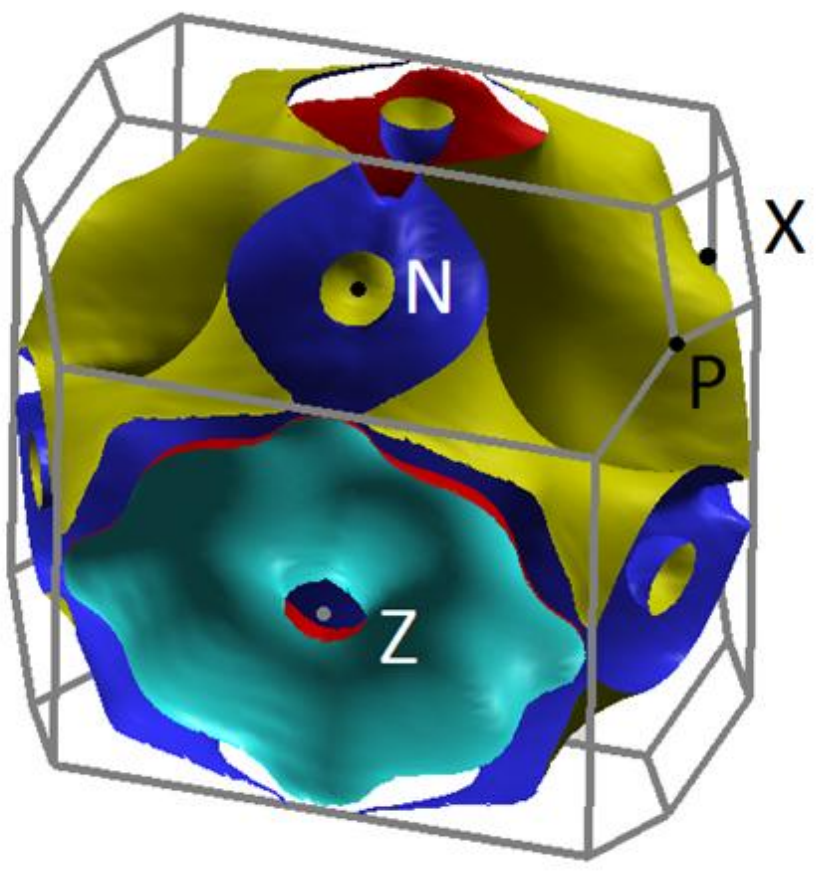

Fermi surface at 3 GPa

5. Change of the bader charge with pressure

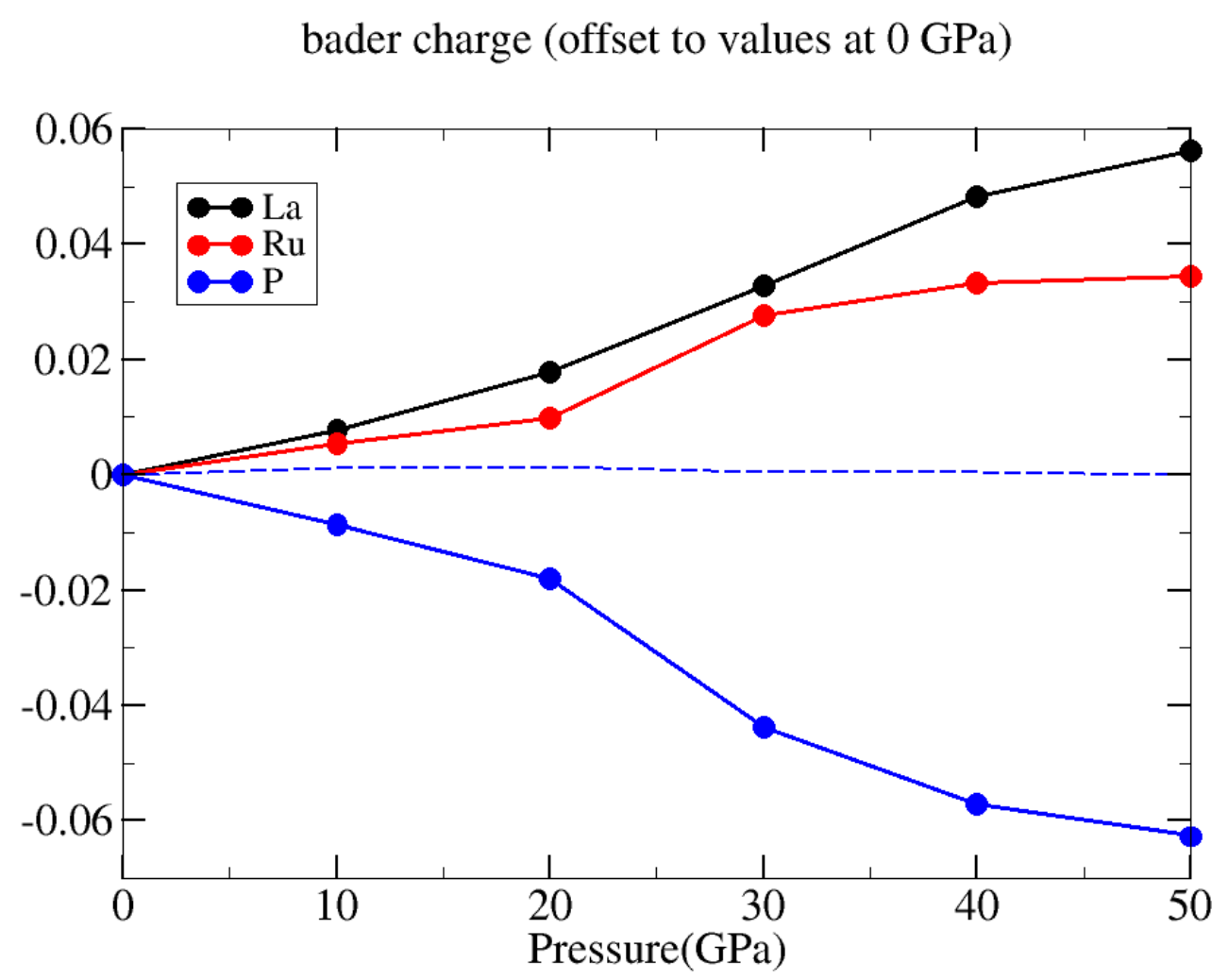

6. Change of the vibrational modes with pressure

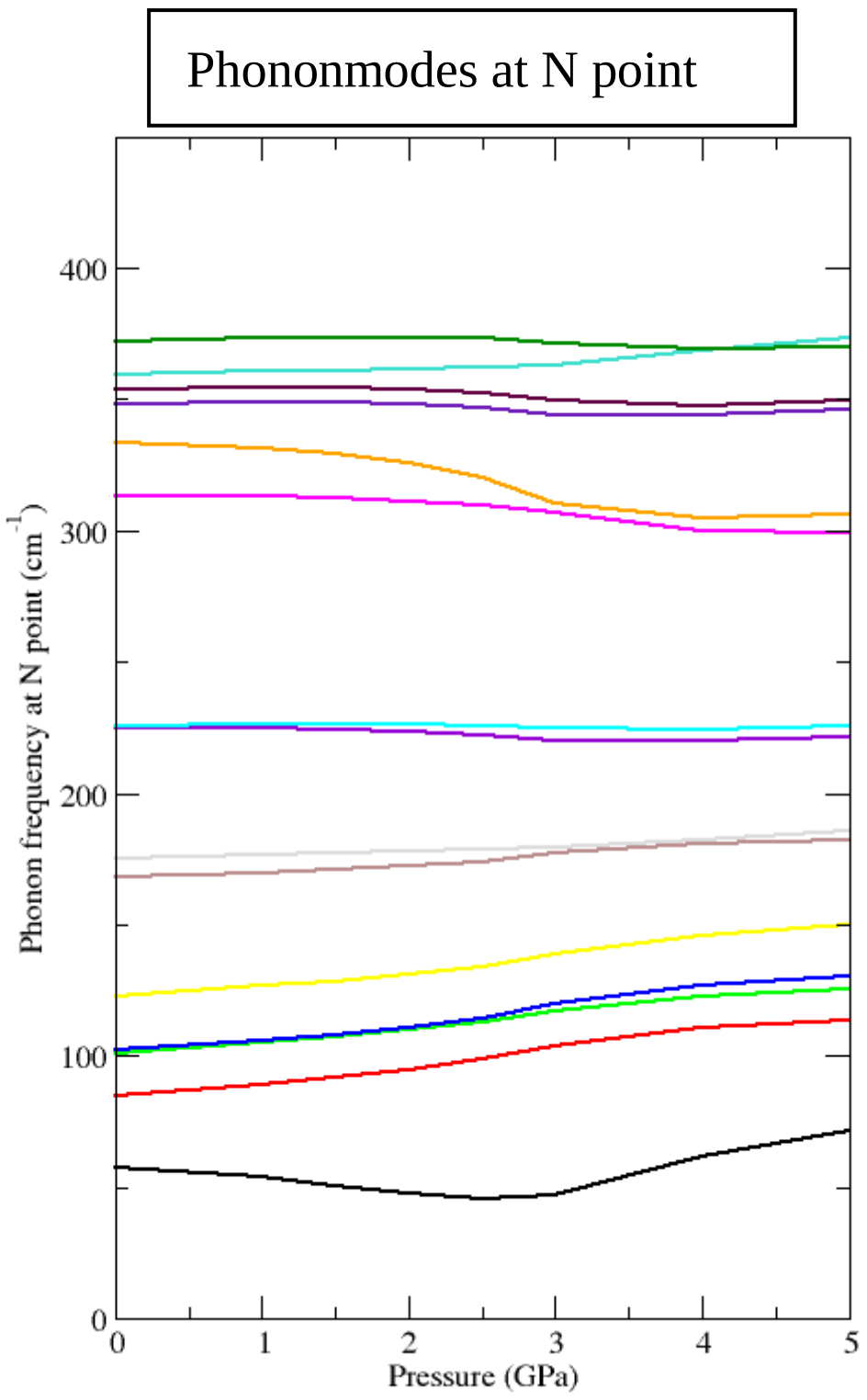